\documentclass[a4paper,11pt]{article}
\pdfoutput=1 % if your are submitting a pdflatex (i.e. if you have
\usepackage{jcappub} % for details on the use of the package, please
\usepackage[T1]{fontenc} % if needed
\usepackage{indentfirst}
\usepackage{comment}

\title{\boldmath Effective field theory for compact object evolution in binary inspirals}

\author[]{Irvin Martínez}
\author[]{\& Amanda Weltman}
\affiliation[]{High Energy Physics, Cosmology \& Astrophysics Theory Group, Department of Mathematics \& Applied Mathematics, University of Cape Town, Cape Town, 7701, South Africa}

\emailAdd{mrtirv001@myuct.ac.za}
\abstract{Using the effective field theory framework for extended objects we describe the evolution of spinning compact objects in the late inspiral of the coalescence of a binary, before the plunge and merger, by including leading order corrections due to spin, tides, dissipation and gravitational wave radiation. Our implementation is of particular relevance for probing the stellar structure of compact objects with gravitational wave observations. A spinning compact object in the effective field theory framework is described as a spinning point particle, with its finite size effects encoded in higher order operators in the effective action, operators which have coefficients that encapsulates the internal structure of the star. For the inspiral regime described by non-relativistic general relativity, post-Newtonian corrections to each term of the action can be obtained in a diagrammatic approach, including gravitational radiation effects. Taking into account the aforementioned ingredients of the effective theory, we solve for the dynamics of the inspiral of binary systems using an algorithm for point particle simulations. We extract the gravitational wave as a function of the orbital frequency, input that is generated numerically and then evaluated in the analytic function of the waveform. By performing illustrative numerical experiments of systems that the LIGO-Virgo observatories have already detected, we show the role of the stellar structure and its coefficients in the phase evolution of the waveform, as well as the order in which they arise and the sensitivity required for the gravitational wave observatories to measure them. If these coefficients are to be measured, tight constraints on fundamental physics can be placed.}

\begin{document}
\maketitle
\flushbottom

\section{Introduction}
\label{sec:intro}

We live in a time with access to data from the most energetic "particle" collisions in the universe. The astonishing first detection of gravitational waves (GWs) by the LIGO-Virgo observatories from a binary black hole (BBH) merger \cite{Abbott:2016blz},  the multi-messenger detection from a binary neutron star (BNS) collision  \cite{TheLIGOScientific:2017qsa, LIGOScientific:2017ync, LIGOScientific:2017zic}, and the recent detection of the coalescence of black hole (BH) - neutron star (NS) binaries \cite{LIGOScientific:2021qlt}, provide us with ample opportunities to test fundamental physics in the strong regime of gravity. 

These detections of binary systems through GWs have already allowed us to better understand the properties of compact objects \cite{LIGOScientific:2018mvr,Abbott:2020gyp}. With more sensitivity upgrades planned for the LIGO-Virgo observatories, the upcoming third generation Einstein Telescope \cite{Punturo:2010zz,Maggiore:2019uih} and the space based detector, LISA \cite{Barausse:2020rsu}, the era of precision gravity is upon us, and with it the potential for great discovery using multi-messenger astronomy to better understand the universe on all scales. On the theoretical side, the development of improved tools to model the dynamics of any astrophysical source to extract the gravitational wave signature, can only aid in this discovery.

One of the key potentials with GW observations is to test fundamental physics by probing the internal structure of the compact objects. For example, the equation of state of matter for NSs is still unknown, and it has been suggested that it can be constrained by matching the Love numbers of the NSs with GW observations \cite{Flanagan:2007ix, Hinderer:2007mb}. The Love numbers are dimensionless parameters that measure the rigidity and tidal deformation of the stellar object, and different values for the Love numbers correspond to different equations of state of matter \cite{Yagi:2013awa}. On the other hand, BHs within the theory of general relativity are well constrained by the no hair theorem, which states that a BH can be described by only three parameters, its mass, spin and charge, which implies the vanishing of its Love numbers \cite{Poisson:2004cw, Chia:2020yla}. Any finding of deviation of this description or discovery of a new parameter that describes the BH must show hints of a more complete underlying gravitational theory \cite{Heisenberg:2018vsk, Giddings:2019ujs}.

Therefore, the modelling of compact objects and their interactions must take into account different effects, such as the spin and the stellar structure of the various stellar types. In this work we bring the tools of effective field theory (EFT) for extended objects \cite{Goldberger:2004jt, Goldberger:2005cd, Delacretaz:2014oxa} to model BHs and NSs as point particles, with additional effects encoded as higher order corrections in the action. Although at first it might seem counter intuitive to model such massive objects as point particles, there is in fact, a lot of similarity in the description of compact objects and elementary particles \cite{Arkani-Hamed:2019ymq}. For instance, the no hair theorem implies that a BH effectively behaves as a point particle, and it has been shown that a spinning BH is described relativistically to all orders in the multipole expansion from minimal coupling, matching on to the effective action of a spinning point particle \cite{Guevara:2018wpp, Chung:2018kqs}.

We start with an effective theory for spinning extended objects derived from the coset construction \cite{Delacretaz:2014oxa, Martinez:2021mkl}, a general technique from the EFT framework that allows us to construct effective actions \cite{Coleman:1969sm}. In this approach, the coefficients appearing in the effective action are treated as free parameters to be fixed by observations. This theory is of particular relevance because it leads to a more natural description of spinning extended objects without the need for redundant degrees of freedom, as is the case for the EFT for spinning extended objects used in \cite{Porto:2005ac, Levi:2008nh}. 

For the stellar structure, we build on \cite{Endlich:2015mke}, where in the context of non-compact objects, static tidal effects and dissipation were considered, using the tools introduced in \cite{Goldberger:2004jt}, and in \cite{Goldberger:2005cd}, for each effect respectively. In this work we go further to consider compact objects and a response function for dynamical tides of NSs derived in \cite{Chakrabarti:2013xza} using the tools from \cite{Goldberger:2004jt, Goldberger:2005cd}, and consider relativistic effects, such as GW radiation, determined by the PN expansion \cite{Goldberger:2004jt}. Although it has been pointed out that this is a theory for "slowly" spinning rigid objects due to the existence of a rotational frequency at which the theory breaks down \cite{Delacretaz:2014oxa}, our current observations of spinning compact objects \cite{LIGOScientific:2018mvr} suggest that we can safely consider that most of the astrophysical objects spin slowly, and therefore are well described by our effective theory.

\begin{figure}
    \centering
    \includegraphics[width=0.95\textwidth]{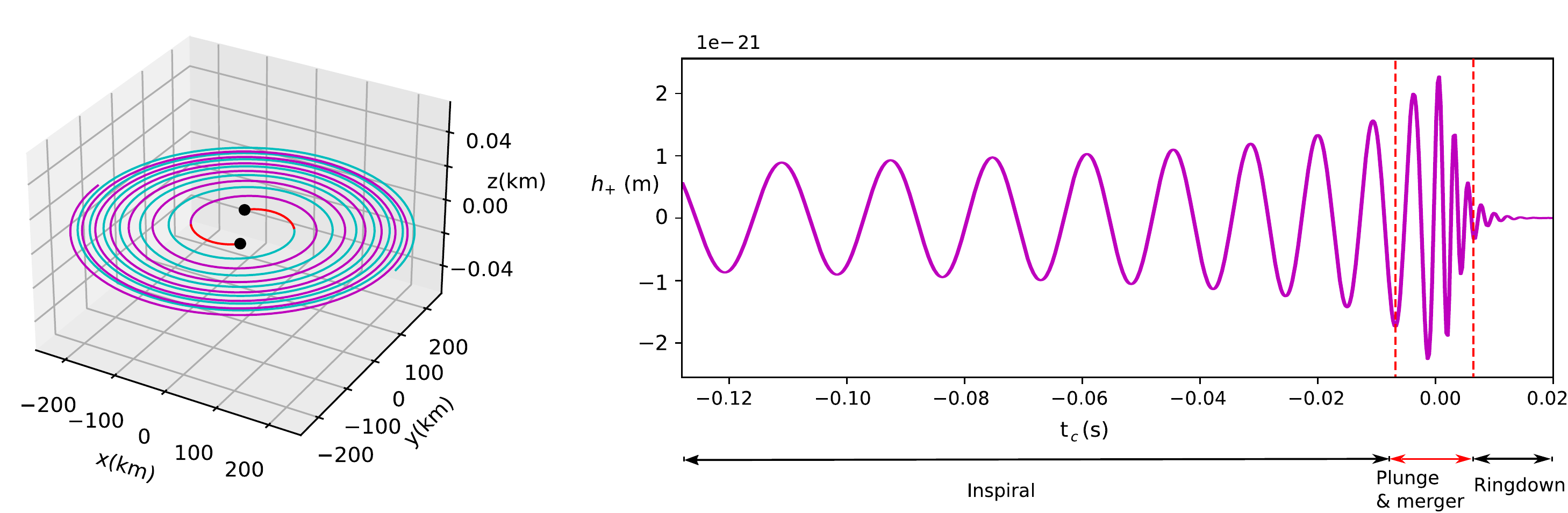}
    \caption{The coalescence of an equal mass BBH system, each with mass, $M = 20 \: M_{\odot}$, and initial gravitational wave frequency, $f_{GW} = 50 \mathrm{Hz}$. The left figure represents the dynamics of the binary until the collision of the BHs using our numerical code for point particles, with the purple and cyan line the trajectories followed by each of the BHs. The orbital trajectory once the LSO is reached is shown in red until the radii of the holes interact. The system decays due to GW radiation by including the leading order 2.5 PN correction in the equations of motion. The figure on the right is the gravitational waveform extracted from the merger of the same binary but using the EOB method from the LIGO Algorithm Library  \cite{lalsuite}. The coalescence is divided into four different phases: inspiral, plunge, merger and ringdown. }
    \label{fig:inspiralGW}
\end{figure}

The description of the coalescence of binary systems through GW radiation is divided into four stages: the inspiral, plunge, merger and ringdown, which is depicted in fig. \ref{fig:inspiralGW}. The inspiral phase is described using the PN expansion \cite{ Goldberger:2004jt,Blanchet:2006zz}. The transition from the inspiral to the plunge can be modelled analytically \cite{Buonanno:2000ef}, although it is usually modelled altogether with the merger by solving the full Einstein's equations numerically \cite{Pretorius:2005gq}, which is computationally expensive. After merging, if the final outcome is a BH, it enters into a ringdown phase that can be modelled analytically using perturbation theory \cite{Leaver:1985ax}.

Furthermore, the need to extract efficient and accurate waveforms for all of the phases of the coalescence led to the development of the effective one body (EOB) framework \cite{Buonanno:1998gg}, which is a combination of analytical methods based on the Hamiltonian, that uses the PN expansion as an input for the inspiral, and non-perturbative methods and numerical relativity results to model the merger semi-analytically. Since the introduction of the EOB many additional improvements have been done to include the stellar structure, such as tidal effects \cite{Damour:2009wj, Hinderer:2016eia, Steinhoff:2021dsn} for NSs, and BHs horizon GW absorption or dissipative effects \cite{Bernuzzi:2012ku}. The EOB is the current framework used for gravitational wave extraction and comparison to observations. In fig. \ref{fig:inspiralGW}, we show a waveform from an equal mass BBH extracted using the EOB method from the LIGO library \cite{lalsuite}.

The properties of the compact objects play an important role mostly in the late inspiral phase of the coalescence, and thus we restrict ourselves to this regime where the dynamics are described by the PN approximation, before the plunge and merger of the binary. The PN expansion encodes non-linearities of the theory of general relativity and its predictability has been tested by comparing PN to numerical relativity simulations. It has been found that the PN expansion reproduces the waveform with high accuracy up to a few orbits before the merger \cite{Baker:2006ha,Boyle:2007ft},  where the PN approximation becomes less accurate as one approaches the inner most circular orbit, or last stable orbit (LSO)  \cite{PhysRevD.80.084043}. The LSO is the closest distance at which a point particle has a stable orbit around the stellar object and it will set the limit of our theory. Once the LSO has been reached, the binary enters into the plunge phase.  In figure \ref{fig:inspiralGW}, we show the dynamics of the coalescence of an equal mass BBH in a quasi-circular orbit made with our simulations, and show where our theory breaks down.

The PN expansion in the EFT for extended objects can be obtained by computing the corresponding Feynman diagrams of the interaction \cite{Goldberger:2004jt}. In this approach, relativistic corrections are described as a perturbative series in terms of the expansion parameter, $v/c <1$, with $v$ the relative velocity of the binary. For each $n$-PN order, the expansion of the equations of motion is of order $v^{2n}$. Although high order PN corrections are necessary for a precise description of the interaction, in this paper we consider only the leading order PN correction due to gravitational wave radiation, the $2.5-$PN term \cite{Goldberger:2004jt}, and neglect the leading order conservative $1$ and $2$-PN effects which only contribute to a shift of the orbit and do not contribute to the decay of the binary. With the purpose of building an intuitive understanding of our EFT to perform numerical simulations,  we skip the use of Feynman diagrams and take the Newtonian limit of the effective action as in \cite{Endlich:2015mke}, which is contained in the leading order correction of each term in the action. The PN expansion of the effective action is discussed in \cite{Martinez:2022vnx}, where the LO expansion to each of the terms considered in this work is obtained. The state of the art of EFT and the PN expansion can be found in \cite{Goldberger:2004jt,Porto:2005ac,Levi:2008nh, Blanchet:2006zz,Foffa:2011ub,Foffa:2012rn,Porto:2008jj,Porto:2008tb,Levi:2010zu,Porto:2010tr,Levi:2015msa,Porto2016,Levi:2018nxp,Goldberger:2020fot}. 

Taking into account the aforementioned ingredients, and with the point particle approximation in mind, we obtain the equations of motion for the effective action and introduce a known methodology for simulating stellar dynamics as point particles \cite{Hut:1994ys} into the framework of EFT for compact objects.  Our implementation of the coalescence of a binary is tested with analytical results, and the stellar structure effects are internally tested by matching the coefficients of the theory during the simulation. Solving numerically for the point particle as well as for its additional effects, allows us to evolve the system with high accuracy for low computational cost.  We propose a hybrid method for extracting gravitational waveforms as a function of the orbital frequency of the binary, $\omega_s$, input which is generated numerically and then evaluated in the analytical formula of the waveform. Our method reproduces results from known methodologies that evaluates the coalescence time, $t_c$, of the binary.

We perform illustrative numerical "experiments" of the inspiral of the different detected binaries to study the imprints of the stellar structure in the dynamics. With these, we show the role of the stellar structure and its coefficients in the waveform, given that the addition of the stellar properties in the equations of motion generates changes in the orbital frequency, which can be measured as a phase shift in the waveform.  We show the order at which the stellar structure effects arise in the waveform, and the sensitivity that the current LIGO-Virgo observatories will need to detect them. We suggest that the matching of the coefficients due to the internal structure of spinning NSs, $n_{\Omega}$, is the first to be matched from observations, given that it can be the leading order effect, making a wave shift in the waveform of order up to, $\Delta \varphi \simeq O(10)$, given our considered spinning objects. Moreover, we perform a systematic study of the internal structure of the objects, and show that, at least at this level of the implementation, in order to constrain the coefficients of all the leading order effects due to the internal structure of the compact objects, current and future GW observatories need to measure wave shifts with an accuracy of at least, $\Delta \varphi \simeq O(10^{-4})$ radians. 

The plausibility for LIGO-Virgo observatories and the Einstein telescope to detect such effects, is of course dependent on the frequency where the dephasing is accumulated in the signal relative to the most sensitive band of the detectors. At current sensitivity, if the phase shift is of order $O(10^0)$ radians in the most sensitive band, then the effect could be measurable and a systematic study needs to be carried out, i.e. using Fisher matrix estimates \cite{Cutler:1994ys}. With upcoming sensitivity upgrades, a systematic study needs to be done to put bounds on the measurability of such effects, and whether all the effects due to the internal structure of the compact object can be constrained. 

In section \ref{sec:EFT}, we explain the effective action which systematically includes spin/size, finite-size and dissipative effects. Then take the Newtonian limit, derive the equations of motion and consider leading order GW radiation effects to the point mass term of the action. In section \ref{sec:waveform}, we introduce the hybrid method for gravitational wave extraction and the signatures due to the stellar structure. In section \ref{sec:simulations}, we implement the formalism numerically, test it with analytical results and measure the coefficients of the theory as an internal tests of the implementation. We then perform illustrative numerical experiments of the detected binaries by LIGO and quantify the effects in the gravitational waveform. In section \ref{sec:discussion}, we discuss the implications of our results. 

\section{Effective field theory for compact objects}
\label{sec:EFT}

Within the framework of EFT for extended objects \cite{Goldberger:2004jt,Goldberger:2005cd}, we build an effective action for spinning compact objects that takes into account the stellar structure. The key idea of modeling compact objects with EFT is that we can treat them as point particles, with the stellar structure encoded in higher order operators. This is represented by the effective action

\begin{flalign}
\mathcal{S} = \int \mathrm{d} \tau \left\{-mc^2 + \sum_n c_n  \tilde{\mathcal{O}}_n \right\},  
\end{flalign}

\noindent with the first term describing a point particle, and the second, the sum over all possible higher order corrections. The coefficients, $c_n$, are the Wilson coefficients of the effective theory that are to be matched from experiments or from the full known theory, and $\mathcal{O}_n$, higher order operators that are allowed by the symmetries. This sum is in principle infinite, and can be cut-off at the desired accuracy. Some of the Wilson coefficients are related to the Love numbers, making the EFT a very good framework to work with. We refer to the Wilson coefficients that contain Love numbers as Love coefficients. Furthermore, not all the coefficients are Love coefficients, i.e. they instead can encode the capacity of a BH to absorb GWs, or the  viscosity of a NS which can generate an energy loss during the interaction. We simply refer to this coefficient as the dissipation coefficient.

The action for spinning objects in gravity derived from the coset construction \cite{Delacretaz:2014oxa}, taking into account the stellar structure \cite{Endlich:2015mke,Martinez:2021mkl}, has the schematic form

\begin{equation}
    \mathcal{S} = \mathcal{S}_{PP} + \mathcal{S}_{\Omega} + \mathcal{S}_{Q} + \mathcal{S}_{\mathcal{D}}, 
\end{equation}

\noindent where, $\mathcal{S}_{PP}$, is the action for a spinning point particle, $\mathcal{S}_{\Omega}$, gravitational corrections due to spin, $\mathcal{S}_{Q}$, dynamical tides and $\mathcal{S}_{\mathcal{D}}$, dissipative effects. 
We consider each of the leading order terms for these effects in turn, to see their overall role. Thus, the effective action reads

\begin{flalign}
\begin{split}
    \mathcal{S} = &\int \mathrm{d} \tau \left\{ -mc^2 + \frac{I}{2} \tilde{\Omega}_{ij} \tilde{\Omega}^{ij}  + n_{\Omega} \tilde{\Omega}^{i} \tilde{\Omega}^{j} \tilde{C}_{0i0j} + \tilde{Q}^{ij} \tilde{C}_{0i0j} + \tilde{\mathcal{D}}^{ij} \tilde{C}_{0i0j} + .\:.\:. \right\}, \\
\end{split}    
\label{eq:action}
\end{flalign}

\noindent where $\tau$ is the proper time. The indices $i,j,k,...$, denote spatial components of the Lorentz indices and the tildes indicates quantities in the comoving frame. This action is Lorentz invariant and is compatible with all possible symmetries. 

The first two terms of equation (\ref{eq:action}) describes the leading order corrections for a relativistic spinning particle with mass, $m$, moment of inertia, $I$, and angular velocity, $\tilde{\Omega}_{ij} = \Lambda^{\;\;k}_{i} D_{\tau} \Lambda_{kj} $, and $\tilde{\Omega}_{ij} = \epsilon_{ijk} \tilde{\Omega}^{k}$. The third term is a spin correction due to gravity, with $\tilde{C}_{0i0j}$, the traceless Weyl tensor. The fourth term accounts for dynamical quadrupolar tidal effects with, $\tilde{Q}_E = \tilde{F}(\tau) \tilde{C}_{0i0j}$ and $\tilde{F}(\tau)$, a dynamical response function whose Fourier transformation can be Taylor expanded around small frequencies, $\omega$, such that $\mathcal{F}(\tilde{F}) \propto n_E + n'_E \omega^2$.\footnote{We have explicitly separated dissipative effects from the response function in the action compared to \cite{Chakrabarti:2013xza}.} \cite{Chakrabarti:2013xza} The coefficient, $n_E$, is the dimensionful Love number for static tides \cite{Chakrabarti:2013xza}, or simply Love coefficient, which is related to the dimensionless Love number $k$, via $n_E = 2k \ell^5/3G$ \cite{Flanagan:2007ix, Binnington:2009bb}, with $\ell$, the radius of the star. Furthermore the coefficient, $ n'_E$, is the dynamical Love coefficient. These quadrupolar effects encode the fact that the object has finite size. Finally, the last term contains dissipative effects  due to the internal structure of the stellar object, with $\tilde{\mathcal{D}}^{ij}$ a composite operator that represents these leading order additional degrees of freedom in a model independent way \cite{Goldberger:2005cd}. The ellipsis represents the tower of higher order operators not taken into account in this work.

We can ensure that our theory will be predictive by expanding our action in small parameters $v/c$, $\ell/r$ and $\Omega/\Omega_0$ to the desired accuracy. The scales, $\ell$, the radius of the object and $\Omega_0$, the typical frequency, do not appear explicitly in the action but determine the characteristic size of the dimensionful coupling constants. The tower of higher order spin terms is under control as long as the rotational velocity is much less than the speed of sound $c_s$ of the material in the star, i.e. $\Omega \ell \ll c_{s}$ \cite{Delacretaz:2014oxa}. For a star, this rotational frequency would be such that the body would undergo large nonlinear stresses and order one distortions, breaking down the theory.

\subsection{Stellar dynamics}

We perform a Lorentz transformation from the frame embedded in the rigid body to the lab frame.  We then take the Newtonian limit of our effective action, which is contained in the lowest order PN expansion of each of the terms. Therefore, in the lab frame, $\tau = t$, $C_{0i0j} \approx \partial_i \partial_j \Phi$, with $\Phi$ the Newtonian potential, and $\Omega^i = \frac{1}{2}\epsilon^{ijk} \mathcal{R}_{jl} \partial_t \mathcal{R}_{k}^{\; l}$, with $\mathcal{R} (\theta^i)$ a rotation matrix, where the $\theta'$s are the Euler angles describing the orientation of the rigid body \cite{Endlich:2015mke,Martinez:2021mkl}. 

\begin{comment}
$\mathrm{d} \tau \approx \mathrm{d}t \sqrt{1 + 2\Phi -\dot{\vec{y}}^2}$ with $\vec{y}$ the location of the star
\end{comment}

The action then reads,

\begin{flalign}
\begin{split}
    \mathcal{S} = &\int \mathrm{d}t \left\{ \frac{m v^2}{2} -m \Phi + \frac{I}{2} \Omega_{ij} \Omega^{ij} + \frac{n_{\Omega}}{2} \Omega^i \Omega^j \partial_{i} \partial_j \Phi \right.\\
    &\left. + \frac{n_{E}}{4} \partial^i \partial^j \Phi \partial_i \partial_j \Phi  + \frac{n'_{E}}{4} \partial^i \partial^j \dot{\Phi} \partial_i \partial_j \dot{\Phi} + \frac{1}{2} \partial_i \partial_j \Phi \mathcal{R}^{i}_{\; k} \mathcal{R}^{j}_{\; l} \tilde{\mathcal{D}}^{kl}_{E} + .\;.\;. \right\}.
\end{split}
\label{eq:newtonianaction}
\end{flalign}

\noindent Each term in the action has a simple physical interpretation as described in \cite{Endlich:2015mke}. The first three terms describe a non-relativistic spinning point particle coupled to gravity. The fourth term describes the coupling between gravity and the ensuing quadrupole. By expanding the gravitational potential around some background value $\bar{\Phi}$, this term can be seen as a deformation of the inertia tensor of the form $\delta I \propto n_{\Omega} \partial^2 \bar{\Phi}$. The first term in the second line describes the coupling of the induced quadrupole $\delta Q \propto n_E \partial^2 \bar{\Phi}$, and the second term its dynamical part. The last term contains dissipative effects, which are encoded in the composite operator $\tilde{\mathcal{D}}^{ij}_E$. 

As we are not interested in explicitly keeping the degrees of freedom encoded in $\tilde{\mathcal{D}}^{ij}_E$, it is necessary to average over them in a systematic way using the in-in formalism \cite{Jordan:1986ug} for this context \cite{Goldberger:2004jt, Endlich:2015mke}, which for dissipative systems, leads to obtainment of the equations of motion from a modified variation \cite{Galley:2012hx}, 

\begin{equation}
    \delta \mathcal{S} + i \int \mathrm{d}t  \mathrm{d}t' \delta J_{ij} (t) \tilde{G}^{ijkl}_R (t - t') J_{kl}(t') = 0, 
    \label{eq:modified}
\end{equation}

\noindent where $J_{kl} = \frac{1}{2} \partial_i \partial_j \Phi \mathcal{R}^{i}_{\;k} \mathcal{R}^j_{\;l}$, and $\tilde{G}_R$ is the retarded correlation function of the operators $\tilde{\mathcal{D}}^{ij}_E$ \cite{Endlich:2015mke}. Then, by considering low frequencies, from which we assume that the degrees of freedom from $\tilde{\mathcal{D}}^{ij}_E$  are near equilibrium, the time ordered two point correlation function imply that the Fourier transform $\tilde{G}_R$ must be an odd, analytic function of the frequency $\omega > 0$ \cite{Goldberger:2005cd}. Thus, the retarded correlation function $\tilde{G}_R$ reads

\begin{equation}
    \tilde{G}^{ijkl}_R (\omega) \simeq \eta_{E} \omega \left( \delta^{ik} \delta^{jl} + \delta^{il} \delta^{kj} - \frac{2}{3} \delta^{ij} \delta^{kl} \right),
    \label{eq:retarded}
\end{equation}

\noindent with the coefficient for dissipative effects, $\eta_E \geq 0$. 

Then, by combining eq. (\ref{eq:retarded}) with the  modified variation eq. (\ref{eq:modified}), and varying it with respect to $x^i$ and $\theta^i$, the equation of motion reads

\begin{flalign}
\begin{split}
    m \dot{v}_i = &-m\partial_{i} \Phi + \frac{n_{\Omega}}{2}\Omega^k \Omega^j \partial_i \partial_j \partial_k \Phi \\
     &+  \frac{n_{E}}{2} \partial_{i} \partial_{j} \partial_{k} \Phi \partial^{j} \partial^{k} \Phi + \frac{n'_{E}}{2} \partial_{i} \partial_{j} \partial_{k} \dot{\Phi} \partial^{j} \partial^{k} \dot{\Phi} \\
    &- \frac{\eta_{E}}{2} \partial_{i} \partial_{j} \partial_{k} \Phi (\partial^j \partial^k \dot{\Phi}  + 2 \partial^j \partial_l \Phi \epsilon^{klm} \Omega_m),
    \label{eq:eom}
\end{split}    
\end{flalign}

\noindent for the acceleration of the compact object, and

\begin{flalign}
\begin{split}
    &\partial_{t} (I \Omega_i + n_{\Omega} \Omega^j \partial_i \partial_j \Phi ) = - n_{\Omega} \epsilon_{ijk} \Omega^k \Omega^l \partial^j \partial_l \Phi \\ 
    &+ \eta_{E} \partial^{j} \partial^k \Phi (3 \partial_i \partial_j \Phi \Omega_{k} - 2 \partial_{j} \partial_{k} \Phi \Omega_{i} + \epsilon_{ikl} \partial^l \partial_{j} \dot{\Phi}),
\end{split}
\end{flalign}{}

\noindent for the change of rotational angular momentum with contributions from $n_{\Omega}$ and $\eta_{E}$. These equations of motion were originally obtained in \cite{Endlich:2015mke}, with the exception of the dynamical tides term with coefficient $n'_E$, which we have derived. 

By substituting the Newtonian potential, $\Phi$, and restricting the spin, $\Omega^i$, to be aligned with the angular momentum of the binary, we obtain

\begin{flalign}
\begin{split}
    m_1 \dot{\vec{v}} = &-\frac{G m_1 m_2 \hat{r}}{r^2} - \frac{3 n_{\Omega} G m_2 \Omega^2 \hat{r}}{2r^4}  - \frac{9 n_{E} G^2 m_2^2 \hat{r}}{r^7} 
     \\ &-  \frac{9 \eta_E G^2 m_2^2}{r^8}(\vec{r} \times \vec{\Omega} + \vec{v} + 2 \hat{r}(\hat{r}\cdot \vec{v}))\\
     &- \frac{18 n'_E G^2 m_2^2}{r^9} \left( 2 v^2 \hat{r}  + 5  (\hat{r} \cdot \vec{v})^2 \hat{r} -\vec{v} (  \hat{r} \cdot \vec{v})  \right), \\
\end{split}
\label{eq:acceleration}
\end{flalign}

\noindent and

\begin{flalign}
\begin{split}
    \label{eqn:spinev}
    &\left( I  + \frac{G m_2  n_{\Omega}}{r^3}  \right) \partial_{t} \vec{\Omega}  = \frac{3Gm_2 n_{\Omega} (\hat{r} \cdot \vec{v}) \vec{\Omega}}{r^4} + \frac{9 \eta_E G^2 m_2^2}{r^7}\left(\hat{r} \times \vec{v} - r \vec{\Omega}\right).
\end{split}
\end{flalign}

\noindent These equations of motion are to be integrated numerically. We will denote the acceleration of the star as $\vec{a}_{\star} = \vec{a}_{PP} + \vec{a}_{\Omega} + \vec{a}_{Q} + \vec{a}_{\mathcal{D}}$, with $a_{PP}$, the point particle or Newtonian term, $a_{\Omega}$, the gravitational correction due to spin, $a_Q$, tides both static and dynamic, and $a_{\mathcal{D}} $ dissipative effects. Although we have restricted the spin to be aligned for simplicity, in principle any spin can be considered. Furthermore, these equations of motion are valid for any stellar object that can be described by an equation of state of matter, with the only difference on the values that the coefficients will take for different stellar types. 

\subsection*{Slowly spinning objects}

As previously mentioned, our theory is valid as long as, $\Omega \ell \ll c_{s}$ holds, which can be tested once we specify the properties of the objects below. First, it is important to note that instead of specifying values of the angular velocity, we will work in terms of the dimensionless spin, $\vec{\chi}$, of the object 

\begin{equation}
    \chi = \frac{c J}{G M^2} =  \frac{c I \Omega}{G M^2},
    \label{eq:chi}
\end{equation}

\noindent where, $J = I \Omega$, is the scalar value of the angular momentum of the star. In order for our theory to work appropriately, we need $\chi$, to be slow and non-negligible $ 0 < \chi \ll 1$, such that we have a significant contribution in the simulations and not break down the theory.    

\subsubsection*{Neutron stars}

We start by introducing the inner most circular orbit distance or LSO distance $d_{LSO}$ for a NS. For a non-spinning NS, the spacetime outside it is described by the Schwarzschild metric, which leads to $d_{LSO} = 6Gm/c^2$, with $m$ the mass of the star. For the case of a spinning NS, if the orbiting object is in prograde motion, to first order in $\chi$, the LSO is $d_{LSO, \Omega} = 6Gm/c^2 (1 - 0.54433 \chi)$ \cite{Miller:1998gr}, with $\chi$ being the dimensionless spin parameter defined above. Nevertheless, for our simulations we use the $d_{LSO}$ for the non-spinning star as the limit of the simulation. The moment of inertia for a NS, can be approximated as \cite{Pethick}

\begin{equation}
    I = 0.21 \frac{m \ell^2}{1 - 2Gm/\ell c^2},
\end{equation}

\noindent where $\ell$ is the radius of the NS. We take the speed of sound inside the star $c_s \lesssim c/\sqrt{3}$ \cite{Bedaque:2014sqa}. 

To take into account tidal effects, we use the static Love coefficient, $n_E =  2 k \ell^5/3G$, with $k = [0.449154,0.36260,0.259909]$, the dimensionless quadrupolar ($k = k_2$) Love number for NSs, for each polytropic index, $n =[0.5,0.7,1]$, respectively \cite{Flanagan:2007ix, Hinderer:2007mb, poisson_will_2014} and the spin coefficient, $n_{\Omega}$, in the Newtonian limit $ n_{\Omega} = n_{E}$ \cite{Yagi:2016bkt}. For the dynamical component of the tidal effects, considering a NS with $m = 1.2 M_{\odot}$, $\ell = 8.89$ km and polytropic index, $n = 1$, the dimensionless dynamical Love number $k'$, related to $n'_{E}$, is obtained from the response function  of the NS in terms of the frequency $\omega_f$, and dimensionful overlap integral, $\mathcal{I}_{f}$, of the fundamental mode of the star \cite{Chakrabarti:2013xza}, which describes to which extent an external field excites the mode.  $\mathcal{I}_{f}$ corresponds to $\mathcal{I}_{02}$ in \cite{Chakrabarti:2013xza},  where the subscript indicates the fundamental mode, $l=0$, and the  quadrupolar moment, $k=k_2$.
The response function reads \cite{Chakrabarti:2013xza}

\begin{flalign}
\begin{split}
    \mathcal{F}(F) & = \frac{1}{2} \frac{\ell^5 }{G}  \frac{q_f^2}{\ell^2 (\omega_{f}^2 - \omega^2)/c^2} = \frac{1}{2} \frac{\ell^5}{G}  \frac{q_f^2}{\ell^2 \omega_f^2/c^2} \left( 1 + \frac{\omega^2}{\omega_{f}^2} + .\;.\;.\right) ,
\end{split}    
\end{flalign}

\noindent where we have expanded over $\omega/\omega_{f}$, and $q_{f}$, is the dimensionless overlap integral which is related to the $\mathcal{I}_f$ through $q_{f}^2 = G \mathcal{I}^2_{f} / \ell^3$. The combination, $\ell \omega_f/c$, is dimensionless as well. 

We identify the Love coefficient

\begin{equation}
    n_E = \frac{1}{2} \frac{ \ell^5}{G}  \frac{q_{f}^2}{\ell^2 \omega^2_{f}/c^2},
\end{equation}

\noindent and comparing to the Newtonian tidal Love number, $n_E =  2 k \ell^5/3G$, we identify the dimensionless Love number $k$ 

\begin{equation}
    k = \frac{3}{4} \frac{q_{f}^2}{\ell^2 \omega^2_{f}/c^2},
\end{equation}

\noindent from which we find agreement with $k$ obtained using the Clairaut-Radau equation \cite{poisson2014gravity}. From the above expansion of the response function, we can extract the term $n'_{E}$

\begin{equation}
    n'_{E} = \frac{1}{2}\frac{\ell^7}{G c^2} \frac{q_{f}^2}{\ell^4 \omega_{f}^4 /c^4} = \frac{1}{2}\frac{\ell^7}{G c^2} k' ,
\end{equation}

\noindent for a limit in which $\omega/\omega_f \ll 1$. We have defined $k'$ as the dimensionless part of $n'_E$ and found $k' = 1.1311$, for this particular model of NS with polytropic index $1$. 

For all the considered NS simulations, we consider the mass and radius defined above, even when considering different equations of state. From the considered speed of sound inside the NS, and its radius, we can safely add spin, $\chi \leq 0.2$, without breaking down the theory. Furthermore, is worth noting that the coefficient regarding dissipative effects in NSs is unknown and is to be matched from hydrodynamical numerical simulations.

\subsubsection*{Black holes}

 The radius of a Schwarzschild BH is $\ell_{\bullet} = 2GM/c^2$. The LSO distance for a nonspinning BH is $d_{LSO} = 6Gm/c^2$. The Love numbers for BHs within general relativity are expected to be zero \cite{ Chia:2020yla,Binnington:2009bb}. The extra parameter that can describe a black hole besides its mass and charge, encodes the capacity for the BH to absorb GWs through its horizon.
Dissipative effects for BHs due to GW radiation in a binary decay were first taken into account in \cite{Poisson:2004cw}, and added into the EFT framework in \cite{Goldberger:2005cd}. In the latter, the graviton absorption cross section for nonrotating black holes was matched from \cite{Starobinskil:1974nkd,Page:1976ki}, in which the response function containing dissipative effects was obtained, and from which the dissipative coefficient can be read off. The coefficient for dissipative effects for nonrotating BHs reads \cite{Goldberger:2005cd} 

\begin{equation}
    \eta_{E} = \frac{16}{90} \frac{G^5M^6}{c^{13}} = \frac{\ell^6_{\bullet}}{360 G c},
\end{equation}

\noindent with $M$ the mass of the BH.

In the spinning case, there are more coefficients to account for dissipative effects, including the coefficient, $\eta_{E}$, for the nonrotating black hole \cite{Chia:2020yla}. The explicit expression of these coefficients has been shown in \cite{Martinez:2021mkl}. Given that our current numerical implementation does not allow us to take into account these extra coefficients, we do not perform simulations of rotating BHs.

\subsection{Non-relativistic general relativity}

In the EFT for extended objects \cite{Goldberger:2004jt}, the PN expansion for the conservative and radiative dynamics of the binary is organized and obtained by solving tree level Feynman diagrams. This framework has allowed us to obtain higher order PN corrections than those derived using different methods, reproducing the well known results up to 3-PN order \cite{Goldberger:2004jt, Foffa:2011ub}, and deriving new results for the 4-PN \cite{Foffa:2011ub} and recently for the 5-PN \cite{Foffa:2019hrb} corrections for the non-spinning case. Spinning compact objects were introduced into the EFT in \cite{Porto:2005ac} to obtain the PN contributions from the spin-orbit and spin-spin coupling, and further corrections were obtained in \cite{Levi:2008nh, Porto:2008jj, Porto:2008tb, Levi:2010zu, Porto:2010tr, Levi:2015msa,Goldberger:2020fot}. The state of the art for spinning objects is 4-PN order \cite{Levi:2016ofk}. A review on the PN expansion using EFTs can be found in \cite{Porto2016,Levi:2018nxp}.

Our model for spinning objects has a different construction \cite{Delacretaz:2014oxa,Martinez:2021mkl}, for which the leading order PN expansion has been obtained in \cite{Martinez:2022vnx}, reproducing known results from \cite{Levi:2015msa}. The Newtonian action, eq. (\ref{eq:newtonianaction}), is contained in the leading order PN correction to each effect \cite{Martinez:2022vnx}. Nevertheless, our work focuses in the internal structure of the compact objects, and the implementation of the conservative PN corrections is left for a future work. The leading order non-conservative correction to the point particle due to gravitational wave radiation, can be extracted from the literature \cite{Goldberger:2004jt, Blanchet:2006zz}.

\subsubsection*{Post-Newtonian dynamics}

From the EFT perspective, we can obtain the PN corrections to each of the terms in the action, eq. (\ref{eq:newtonianaction}). We make use of the PN expansion to the point particle term \cite{Goldberger:2004jt}.
The PN corrections to the point mass can be expanded in a series as

\begin{equation}
    \vec{a}_{PN} = \vec{a}_{PP} + c^{-2} \vec{a}_2 + c^{-4} \vec{a}_4 + c^{-5} \vec{a}_5 + O(c^{-6}),
\end{equation}

\noindent where, $\vec{a}_{PP} = \vec{a}_{N}$, is the point particle acceleration or Newtonian term. The terms $\vec{a}_2$ and $\vec{a}_4$ are the first and second order PN contributions of the conservative sector which accounts for, i.e. the periastron shift. The leading order term that accounts for radiation of energy and momentum through GW radiation is the 2.5 PN term, $\vec{a}_{5}$, which we denote as, $\vec{a}_{GW} = \vec{a}_{5}$. This correction reads \cite{Goldberger:2004jt, Blanchet:2006zz}

\begin{flalign}
\begin{split}
    \vec{a}_{GW} = &\frac{4}{5} \frac{G^2 m_1 m_2}{r^{3}} \left[ \left(\frac{2Gm_1}{r} - \frac{8Gm_2}{r} - v^2 \right)\vec{v} +  (\hat{r}\cdot \vec{v}) \left(\frac{52Gm_2}{3r} -\frac{6Gm_1}{r} + 3v^2 \right)\hat{r}\right].
    \label{eq:gwdecay}
\end{split}    
\end{flalign}

\noindent For the rest of the paper we  make use of the modified acceleration, $\vec{a} = \vec{a}_{\star} + c^{-5}\vec{a}_{GW}$,  where $ a_{\star}$, is the acceleration which includes the Newtonian term and the stellar structure of the star that we have taken into account in eq. (\ref{eq:acceleration}). We neglect the PN contribution from the conservative sector given that it does not contribute to the decay of the binary. The GW decay due to the $2.5$ PN term is shown in Fig. \ref{fig:inspiralGW} for an equal mass BBH.

\section{Waveforms and observational signatures}
\label{sec:waveform}

\subsection{Waveform extraction}

Gravitational radiation is produced at lowest order by the time varying mass quadrupole moment.  An analytical expression for the quadrupole gravitational waveform from the inspiral of a binary can be obtained in the setting of flat background spacetime with linearized gravity. One can solve the linearized Einstein's equations in the presence of the binary as a matter source and obtain a solution for the amplitude, project it into the transverse-traceless gauge, and expand it as a multipole expansion. The solution is the amplitude for each polarization mode, $h_{+}$ and $h_{\times}$ \cite{Maggiore:1900zz}.

Binaries whose orbital distance is large enough, such that changes in the orbital distance due to GW radiation over several periods are small, can be considered as binaries with fixed orbits. In this scenario, the quadrupolar, $+$ polarized, GW amplitude for an inspiral binary in a circular fixed orbit reads \cite{Maggiore:1900zz} 

\begin{flalign}
h_{+} (\omega_s, t) = \frac{2}{r_o} \frac{ G^{5/3}}{c^4} \frac{m_1 m_2}{m^{1/3}}  \omega_s^{2/3}  \cos (2 \omega_s t) = \frac{2 G \mu x}{c^2 r_o} \cos (2 \omega_s t),
\label{eq:amplitudefixed}
\end{flalign}

\noindent where $r_o$, is the distance from the observer to the binary, $\mu$, the reduced mass, $m$, the total mass of the binary, and $\omega_s$, the orbital frequency. We have introduced the dimensionless variable, $x \equiv \left( G m \omega_{s}/c^3 \right)^{2/3}$. Furthermore, we have chosen the inclination angle, $\iota = \pi/2$, such that, the $\times$ polarized amplitude, $h_{\times}$, vanishes. 

 For the late inspiral regime, where the changes in the orbital distance and frequency are relevant, it is necessary to take into account the orbital decay in the waveform. This can be done by obtaining the radiated power, the amount of energy radiated through GWs per unit time, and relating it to the change of energy of the system, $\dot{E}_{orb}$, as $P = -\dot{E}_{orb}$. From the latter, an expression for the change of the orbital frequency in time, $\dot{\omega}_s$, is obtained in terms of $\omega_s$. By integrating it over the coalescence time measured by an observer, $t_c = T_c - t$,  with $T_c$, the time that the binary takes to merge, one obtains the relation between orbital frequency and the time of coalescence,

\begin{flalign}
    \omega_s (t_c) =  \frac{1}{8} \left(\frac{5 c^{5}}{G^{5/3}} \frac{m^{1/3}}{m_1 m_2} \right)^{3/8} t_c^{-3/8}.
    \label{eq:freqtau}
\end{flalign}

\noindent Then, in eq. (\ref{eq:amplitudefixed}), one replaces the phase, $2 \omega_s t \rightarrow \varphi (t)$, with the accumulated orbital phase,

\begin{flalign}
    \varphi (t) = 2 \int_{t} \mathrm{d}t \; \omega_s (t) = - 2 \int_{t_c} \mathrm{d}t_c \, \omega_s (t_c),
\end{flalign}

\noindent where we have used $\mathrm{d}t = - \mathrm{d} t_c$. By substituting eq. (\ref{eq:freqtau}) into last equation, solving the integral for the time of coalescence, and expressing it in terms of the orbital frequency, one obtains the accumulated orbital phase,

\begin{flalign}
    \varphi(x) = - \frac{1}{32} \frac{m^{2}}{m_1 m_2} x^{-5/2} + \varphi_0 = - \frac{1}{32} \frac{m^{2}}{m_1 m_2} x^{-5/2} - \frac{2Gm\omega_s}{c^3} \log \left(\frac{\omega_s }{\omega_0} \right),
\end{flalign}

\noindent where $\varphi_0$ has been fixed from  \cite{Blanchet:2006zz}, and with $\omega_0$, a constant frequency that is chosen as the frequency at which the binary enters into the detectable band.

Thus, the gravitational waveform that takes into account the GW decay, reads

\begin{flalign}
 h_{+} (x (\omega_s)) = \frac{2 G \mu x}{c^2 r_o} \cos 2 \varphi(x)
\label{eq:waveformx}.
\end{flalign}

\noindent with $\mu$ the reduced mass. We simply denote, $h_{+} (x (\omega_s)) = h_{+} (\omega_s)$. This expression is the lowest order or $0-$PN waveform in the PN approximation. By writing $x = ((Gm/r)(r \omega_s/c^3) )^{2/3}$, we find its scaling given that, $Gm/r \sim v^2$, and $r \omega_s \sim v$. Thus, every $x$ scales as $x = O \left( v^2/c^2 \right)$, meaning that corrections in powers of $v/c$ can be expressed in powers of $x^{1/2}$. The PN expansion of the $+$ polarized waveform in powers of $x$, reads \cite{Blanchet:2006zz}

\begin{flalign}
\begin{split}
    h_{+} (x) = \frac{2 G \mu x}{c^2 r_0} \left( H^{0}_{+} + x^{1/2} H^{(1/2)}_{+} + x H^{(1)}_{+} + x^{3/2} H^{(3/2)}_{+} + x^{2} H^{(2)}_{+} +  O(c^{-5})  \right),
    \label{eq:amplitudePN}
\end{split}
\end{flalign} 

\noindent with $H^0_+ = \cos 2 \varphi$. The last expression shows the program to extract high order PN waveforms that matches observations. Nevertheless, for the purpose of this paper we consider only the amplitude in eq. (\ref{eq:waveformx}). 

\subsection{Comparison to known results}

Working with the amplitude formula as a function of the orbital frequency, $\omega_s$, we need to generate the input numerically by solving the equations of motion as shown in the numerical section.  This is in contrast to the amplitude as a function of the time of coalescence, $t_c$ measured by the observer, which can be obtained via eq. (\ref{eq:freqtau}) in eq. (\ref{eq:waveformx}). In this case, the waveform can be simply extracted if the coalescence time is known. The analytical formula for the time of coalescence for two point masses orbiting each other in a circular orbit, a good approximation for a binary black hole, reads \cite{1964PhRv..136.1224P}

\begin{equation}
T_c (d) = \frac{5}{256} \frac{d^4 c^5}{G^3 m_1 m_2 (m_1+m_2)},
\label{eq:timecoal}
\end{equation}

\noindent with $d$, the orbital distance of the binary. Nevertheless, the stellar structure changes the merger time, and deriving analytical expressions becomes a challenge. Thus, to measure the imprints of the stellar structure in the late inspiral of the binary, we extract the waveform as a function of the orbital frequency, $\omega_s$.  

\begin{figure}
    \centering
    \includegraphics[width=0.9\textwidth]{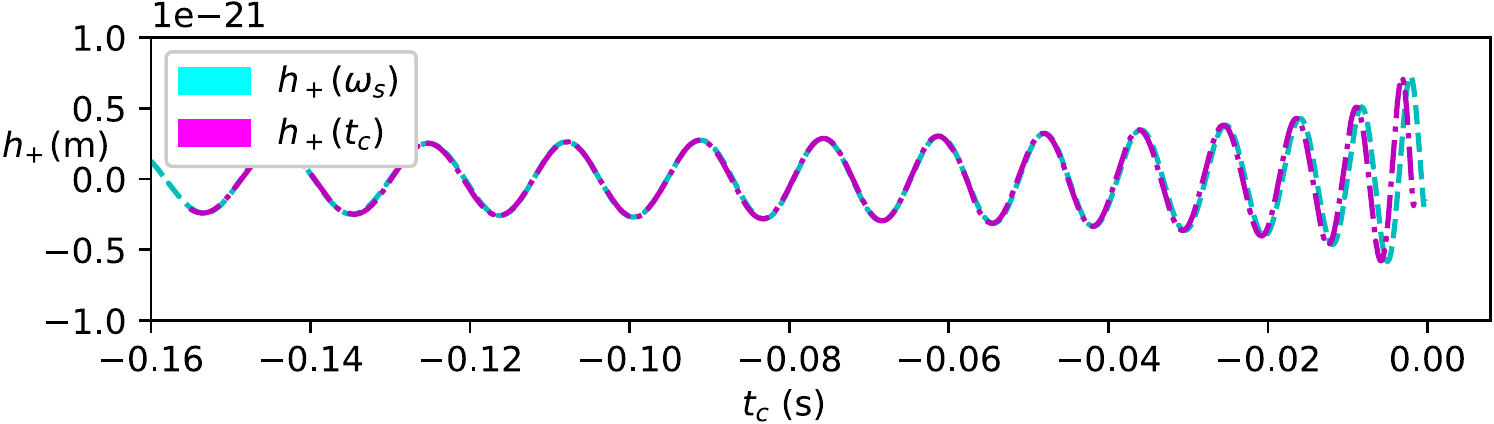}
    \caption{Comparison of the leading order gravitational waveform as a function of the time of coalescence, $t_c$, and of the orbital frequency, $\omega_s$. Using the same equal mass BBH as in figure \ref{fig:inspiralGW}, we extract the orbital frequency and time of coalescence to obtain the waveforms.  The cyan line is the waveform as a function of $\omega_s$, input generated with our numerical simulations. The magenta line is the waveform as a function of the $t_c$, which is obtained from using eq. (\ref{eq:timecoal}) as the only input.}
    \label{fig:comparisonwaveform}
\end{figure}

To compare our methodology with known results, we show the extraction of the orbital frequency from the decay of a binary from the simulations. Consider a binary with masses $m_1$ and $m_2$, and relative distance, $\vec{r} = \vec{r}_2 - \vec{r}_1 $, in some inertial frame.  Then, considering only Newtonian gravity, the relative motion,  $\vec{a} = \vec{a}_2 - \vec{a}_1 $,  reads

\begin{flalign}
    \vec{a}_N = - G \frac{m}{r^2} \hat{r}.
\end{flalign}

\noindent with $m = m_1 + m_2$. The Newtonian orbital frequency, $\omega_N$, is simply obtained from Kepler's law, by equating the centripetal to the Newtonian relative acceleration. The centripetal acceleration, $a_c =  v^2/r$, is a radial force that has the opposite direction to the gravitational radial acceleration. Furthermore, the relative velocity, $v$, is related to the orbital frequency as $v = \omega r$. Thus, the Newtonian orbital frequency is simply given by

\begin{flalign}
    \omega_N = \sqrt{\frac{Gm}{r^3}}.
    \label{eq:orbfreqN}
\end{flalign}

Using the above reasoning, we obtain the orbital frequency with the GW effects included. This can be done by equating the total radial acceleration, including the radial term in (\ref{eq:gwdecay}), to the orbital frequency, $\omega_s$. Thus, the orbital frequency taking into account gravitational radiation, reads

\begin{flalign}
\begin{split}
    \omega_{s} = \omega_{N + GW}
    =& \sqrt{\frac{G m}{r^3} - \frac{4}{5} \frac{G^2 m_1 m_2}{r^{4}} (\hat{r}\cdot \vec{v}) \left(\frac{52Gm}{3r} -\frac{6Gm}{r} + 6v^2 \right)}. \label{eq:GWfreq}   
\end{split}
\end{flalign}

\noindent In figure \ref{fig:comparisonwaveform}, we show the comparison of our method for GW extraction dependent on the orbital frequency using eq. (\ref{eq:waveformx}) and eq. (\ref{eq:GWfreq}),  to leading order in the amplitude as a function of the coalescence time, $t_c$ \cite{Maggiore:1900zz}. Given the initial distance, $d = 10 {\ell}_{\bullet}$, of an equal mass binary, $m_{\bullet} = 20 M_{\odot}$, we can obtain the coalescence time and extract the waveform, $h_{+} (t_c)$ \cite{Maggiore:1900zz}. To extract the waveform as a function of the orbital frequency, $h_{+} (\omega_s)$, we generate  eq. (\ref{eq:GWfreq}) numerically with our code described below, and then evaluate it in eq. (\ref{eq:waveformx}).  Both waveforms match onto most of the inspiral, with some discrepancies near the coalescence time, which might be due to cutting off the PN corrections. Recall that we are only using the leading order GW radiation correction, or 2.5 PN term. With these results, we can safely proceed to study, to leading order, the effects due to the stellar structure.  

\subsection{Observational signatures}

There are observable consequences due to the internal properties of the star that can be quantified. The effective potential changes the acceleration of each object, making the frequency and amplitude of the GW shift. To obtain the frequency shift, we include the stellar structure and derive the orbital frequency in the same way as for the GW decay. We only show the static tidal effects for simplicity, but all other effects are incorporated in the same manner. The contribution from static tidal effects is purely radial, such that the orbital frequency now reads

\begin{flalign}
\begin{split}
    \omega_{N + GW + T}  
    =& \sqrt{\frac{G m}{r^3} - \frac{4}{5} \frac{G^2 m_1 m_2}{r^{4}} (\hat{r}\cdot \vec{v}) \left(\frac{52Gm}{3r} -\frac{6Gm}{r} + 6v^2 \right) + \frac{9 n_{E} G^2 (m_1^2 + m_2^2)}{r^8}} ,    
\end{split}
\end{flalign}

\noindent where the subscript, $T$, in this example, is referring to static tidal effects. 

To quantify the phase shift, $\Delta \varphi$, generated in the waveform (\ref{eq:waveformx}), we compare simulations of binaries including different effects, but the same initial conditions. For instance, consider the orbital frequency from two binaries, one with GW radiation only, $\omega_{N + GW}$, and the other with GW radiation and static tidal effects, $\omega_{N + GW + T}$. From equation (\ref{eq:waveformx}), we find the change in the phase shift due to the additional effects,

\begin{flalign}
    \Delta \varphi = (\varphi_{N+GW+T} - \varphi_{N+GW}).
\end{flalign}

\noindent The phase shift is to be measured at each time step of the simulation.  In general, the phase shift due to the contribution of an additional effect is measured as 

\begin{equation}
    \Delta \varphi = (\varphi_{b} - \varphi_{a}),
\end{equation}

\noindent where the $a$ and $b$ refer to different effects in the simulations.

\section{Numerical simulations}
\label{sec:simulations}

We implement a 4th order Hermite integrator \cite{Hut:1994ys, 1997ApJ...480..432M}.  We briefly review the basics to evolve a system of point particles numerically, in order to solve for the dynamics of the compact objects in a binary system.

\subsection{Point particle simulations}
\subsubsection*{Position, velocity and acceleration}

In the point particle approximation, the predicted values of the position, velocity and acceleration for the next step are obtained from 

\begin{flalign}
\begin{split}
    &\vec{r}_{i+1} = \vec{r}_i + \vec{v}_i \Delta t + \frac{1}{2!} \vec{a}_{i} \Delta t^2 + \frac{1}{3!} \vec{j}_{i} \Delta t^3 + .\;.\;. \, ,\\
    &\vec{v}_{i+1} = \vec{v}_{i} + \vec{a}_{i} \Delta t + \frac{1}{2!} \vec{j}_i \Delta t^2 + .\;.\;. \, ,\\
    &\vec{a}_{i+1} = \vec{a}_{i} + \vec{j}_i \Delta t + .\;.\;. \, ,
\end{split}
\end{flalign}

\noindent where $\vec{j} = \dot{\vec{a}}$. The force calculation is computed as 
\begin{flalign}
    \begin{split}
        &\vec{a}_{i} = \frac{G m_2 \hat{r}}{r^2} + . \;  .  \;  .  \, ,\\
        &\vec{j}_{i} = \frac{G m_2}{r^3} \{\vec{v} - 3(\vec{v}\cdot \hat{r})\hat{r}\} + . \; . \; .  \, ,
    \end{split}
    \label{eqn:acc}
\end{flalign}

\noindent with the position and velocity from the previous step; the ellipses denote all other effects we have taken into account: GW radiation, spin size corrections, dynamical tides and dissipation. The force calculation, $\vec{a}_{i+1}$ and $\vec{j}_{i+1}$, is done using the predicted values, $\vec{r}_{i+1}$ and $\vec{v}_{i+1}$, to get the corrected position, velocity and acceleration

\begin{flalign}
    \begin{split}
        &\vec{r}_{i+1,c} = \vec{r}_{i} + \frac{1}{2}(\vec{v}_i + \vec{v}_{i+1}) \Delta t + \frac{1}{12}(a_i - \vec{a}_{i+1})\Delta t^2 ,\\
        &\vec{v}_{i+1,c} = \vec{v}_{i} + \frac{1}{2}(\vec{a}_i + \vec{a}_{i+1}) \Delta t +  \frac{1}{12}(\vec{j}_i - \vec{j}_{i+1})\Delta t^2 , \\
        &\vec{a}_{i+1,c} = \vec{a}_i + \frac{1}{2}(\vec{j}_i + \vec{j}_{i+1})\Delta t ,
    \end{split}
\end{flalign}

\noindent up to the jerk correction. With this very simple framework we are capable of numerically evolving the equations of motion of our theory, obtaining the desired accuracy by using a specific time-step $\Delta t$.

\subsubsection*{Spin evolution}

We solve for the angular velocity, $\vec{\Omega}$, in a similar fashion as above. We define $\vec{b} = \partial_t \vec{\Omega}$ and correct the angular velocity as follows

\begin{equation}
    \begin{split}   \vec{\Omega}_{i+1,c} = \vec{\Omega}_{i} + \frac{1}{2}(\vec{b}_i + \vec{b}_{i+1}) \Delta t +  \frac{1}{12}(\dot{\vec{b}}_i - \dot{\vec{b}}_{i+1})\Delta t^2 .
    \end{split}
\end{equation}

\subsection{Numerical tests and the matching of coefficients}

\begin{figure*}
\centering
    \includegraphics[width =14.5cm]{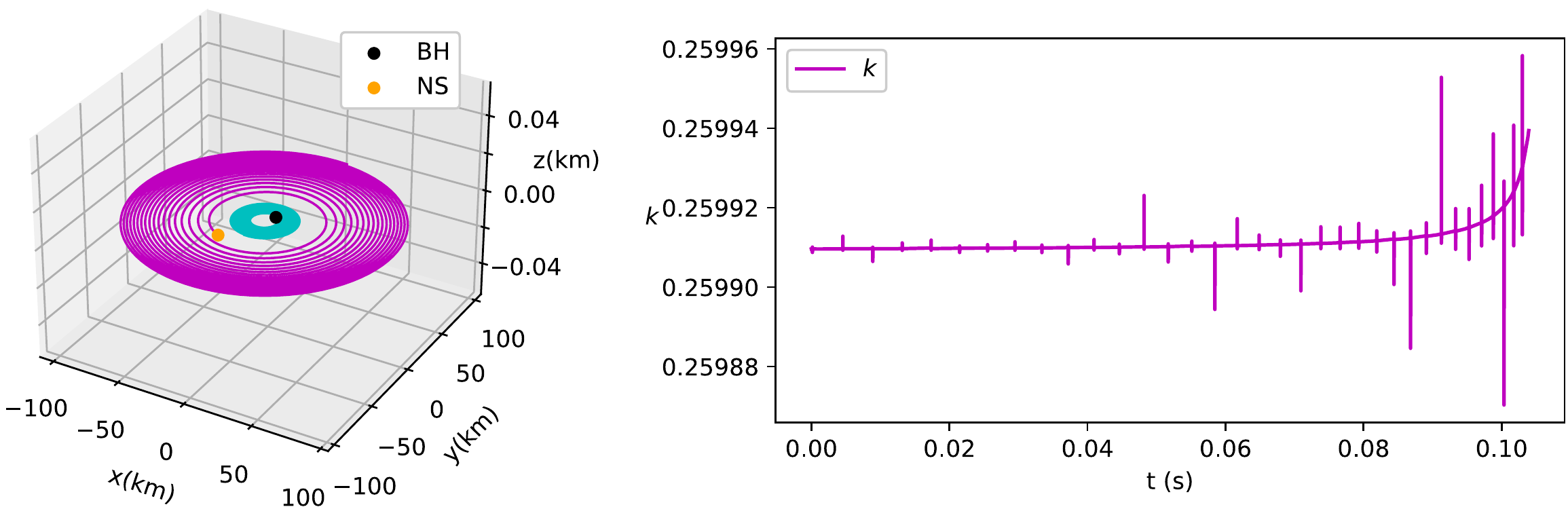}
    
\caption{Simulation of the coalescence of a BH-NS binary and the measurement of the dimensionless static tidal number $k$ of the NS during the interaction. The binary is composed of a BH of mass $m_{\bullet} = 5 M_{\odot}$, and a NS of mass $m_{\star} = 1.2 \: M_{\odot}$ and radius $\ell_{\star} = 8.89$ km. The simulation has an initial orbital distance of $d = 8 \: \ell_{\bullet}$ and it is stopped at the LSO distance of the BH. The figure on the left illustrates the trajectories followed by the binary until the LSO orbit is reached. The purple and cyan line represents the trajectory followed by the NS and the BH respectively. On the right figure the coefficient $k$ is measured at each fixed time-step in the simulation. The theoretical value of the coefficient for the NS used in this case is $k = 2.59909$, which can be measured with high accuracy with an appropriate time step, $dt$. For this simulation,  $dt = 0.1$ in code units. 
}
\label{fig:matching}
\end{figure*}

Many Newtonian and PN codes have been implemented to do numerical experiments.
Testing the Newtonian limit is straightforward. To test the PN corrections we use the GW time decay eq. (\ref{eq:timecoal}). In figure \ref{fig:comparisonwaveform}, we show that our implementation reproduces the leading order waveform as a function of the coalescence time $t_c$. 

To test the rest of the effects we can match the coefficients of the theory from the corrected position, velocity and acceleration at each time-step, which serves as an internal test of the implementation and shows the accuracy of the simulation. From an EFT perspective, we can follow the energy hierarchy to measure different coefficients. Consider for instance a BH-NS binary without any spin, and neglect for a moment the GW radiation effects. The dynamics of the NS will change due to the static tidal effects, and the matching of the dimensionless coefficient reads 

\begin{equation}
    k = \frac{1}{6}\frac{m_1}{m_2 }\bigg(\frac{r}{\ell}\bigg)^{5} \left(\frac{a_{\star, j}}{a_{N,j}} - 1\right),
\end{equation}

\noindent where $a_{N,j}$ is one of the spatial components of the Newtonian term ($j = \{x,y,z\} $) without any other effect, and $a_{\star,j} = a_{N,j} + a_{T,j}$ is the total acceleration of the star, with $a_{T,j}$ being the acceleration due to static tides. Thus, if we solve our system in the $x-y$
plane with these coordinates, we must match coefficients for each spatial component. 
Then we add GW radiation effects and measure the coefficient as

\begin{equation}
    k = \frac{1}{6}\frac{m_1}{m_2 }\bigg(\frac{r}{\ell}\bigg)^{5}  \left(\frac{a_{\star,j}}{a_{N,j}} - \frac{a_{GW,j}}{a_{N,j}} - 1\right).
\end{equation}

\noindent Figure \ref{fig:matching} shows the time evolution measurement of the dimensionless static Love number, $k$, from a BH-NS coalescence. The matching of coefficients from static tidal effects is the simplest, but we can add all other effects and go order by order in the energy hierarchy to measure the coefficients and test the implementation.  

Although it might seem trivial to match the coefficients within the simulation, we suggest that in a similar manner, the matching of coefficients from simulations can be done, to compare, i.e. our state of the art hydrodynamics codes to our EFT for compact objects, a first step towards matching the coefficients from GW observations.

\subsection{Numerical experiments and observations}

The LIGO-Virgo observatories detect gravitational waves from distant astrophysical sources in the frequency range, $f_{GW} \sim [10, 10000]$ Hz \cite{Martynov:2016fzi}, with $f_{GW} = \omega_{GW} / 2 \pi = \omega_s /\pi$ as in \cite{Maggiore:1900zz}, and $\omega_s$ the orbital frequency. Thus, for the purpose of waveform extraction and measurement of the different effects, we run illustrative simulations of the coalescence of various systems with initial and final frequency inside the LIGO-Virgo band. We systematically add the stellar structure effects and extract the lowest order gravitational waveform for different systems, (\ref{eq:waveformx}). We choose an arbitrary distance to the system, $r_o$, that makes the system detectable in the LIGO-Virgo band.  

Each binary system is set in a circular orbit, with initial conditions generated by a simulation with twice the distance of the shown examples. This is done with the purpose of avoiding numerical errors that can be generated in the first orbits of the system. 

\subsubsection*{Neutron star - black hole interactions}

In this interaction, all the effects we have considered play a role in the dynamics: static and dynamical tides in the NS and dissipation in the BH. We divide the BH-NS simulations into two sets, one set to show the role of each effect due to the internal structure of the objects, and the other set to compare the role of different static Love coefficients by considering different equations of state of matter for the NS. In each simulation, we have the same initial conditions of a NS-BH binary, with the mass of the BH, $m_{\bullet} = 5 \: M_{\odot}$, and the mass and radius of the NS, $m_{\star} = 1.2 \: M_{\odot}$ and $\ell_{\star} = 8.89$ km. The simulations start at an initial distance of $d = 25 \: \ell_{\bullet}$ and initial frequency of $f_{GW} = 40$ Hz. The simulations are finished when the NS reaches the $d_{LSO}$ of the BH. Furthermore, we have included slow spin to the NS, $\Omega = [0.1, 0.2]$, aligned to the orbital momentum of the binary.

\begin{figure}
\centering
    \includegraphics[width =0.9\textwidth]{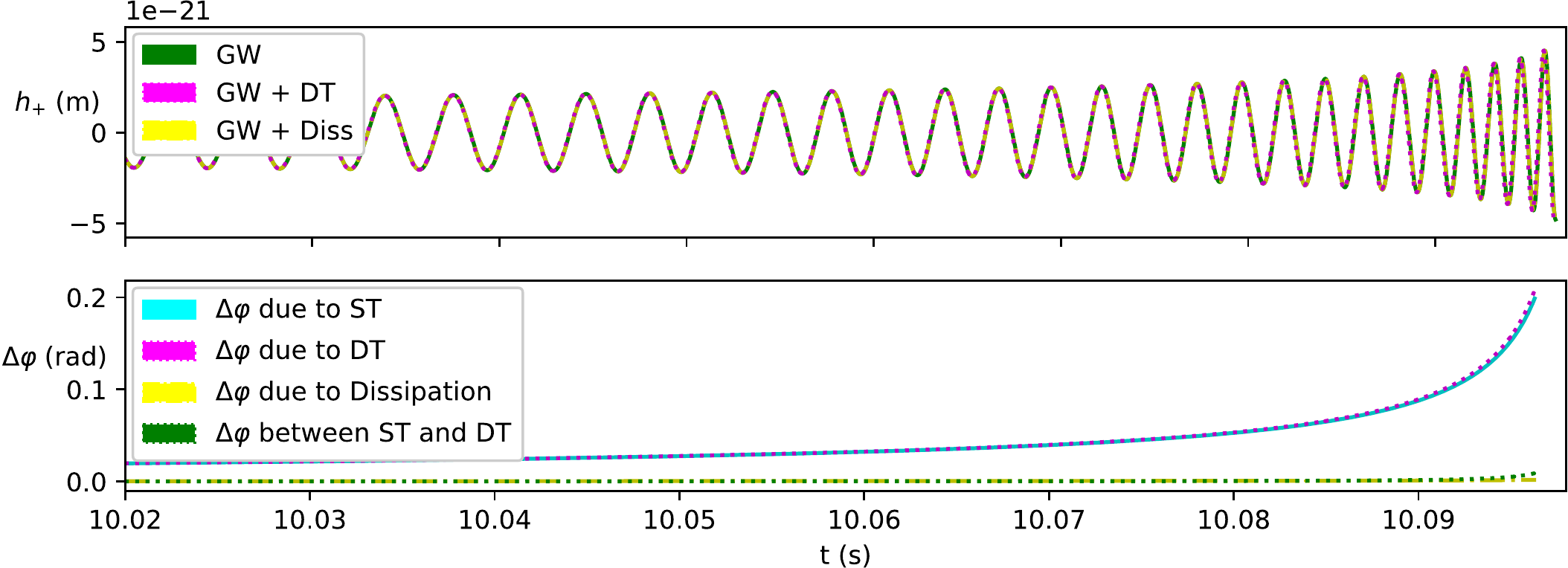}
\caption{Gravitational waveforms and phase shifts from the late inspiral of a BH-NS coalescence with $m_{\bullet} = 5  M_{\odot}$, $m_{\star} = 1.2 \: M_{\odot}$ and $\ell_{\star} = 8.89$ km, at an initial distance of $d = 12 \: \ell_{\bullet}$. The time to the LSO is $T_c \approx 10.096$ sec. We show the very last part of the simulation.  On the top figure, each waveform color is extracted from simulations with different effects. The green waveform is only GWs, the magenta is GWs + dynamical tides (which includes the static part), and the yellow is GWs + dissipation. On the plot below, the phase shift is measured. The cyan line is the wave shift generated by including static tidal effects to the GWs, the magenta line by taking into account static and dynamical tides, the yellow line by dissipation, and the green line is the phase shift that dynamical tides generate with respect to the static part.}
\label{fig:tides}
\end{figure}

First we consider the BH-NS system without any spin. In Fig. \ref{fig:tides}, we compare the phase evolution in the waveforms and quantify the phase shifts from the binaries with different effects:  GWs, GWs $+$ static tides, GWs $+$ dynamical tides, and GWs $+$ dissipation. The effects due to the internal structure are small, as one can not distinguish a difference by eye in the waveforms of the upper plot of fig \ref{fig:tides}. In the lower plot of fig. \ref{fig:tides}, we measure the difference of the accumulated orbital phase, $\Delta \varphi$, due to the different effects. As expected from eq. (\ref{eq:eom}), the major contribution comes from the static tides term, having a final phase shift, $\Delta \varphi_{ST} \simeq 0.2$, in our simulations, which is depicted cyan color. The addition of dynamical tides contributes, $\Delta \varphi_{DT} \simeq 0.009$, with respect to the static tides, which is shown by the green line. Finally, the effects of dissipation from the BH contributes the least, shifting the waveform by $\Delta \varphi_{Diss} \simeq 0.0018$.

\begin{figure}
\centering
    \includegraphics[width =0.9\textwidth]{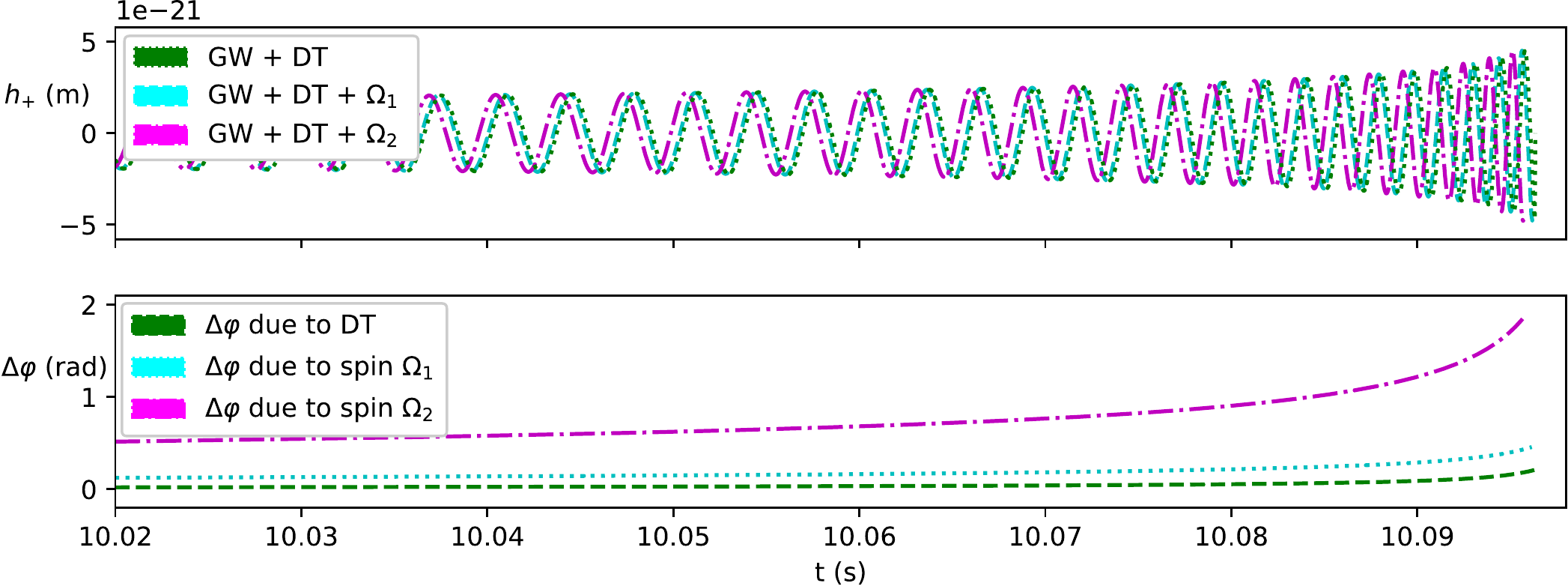}
\caption{Gravitational waveforms and phase shifts from the late inspiral of the same BH-NS interaction described in figure \ref{fig:tides}, but with dimensionless spin $\chi = [0.1, 0.2]$ on the NS. On the top figure, each waveform color is extracted from simulations with different effects. The green waveform is GWs + DT, the cyan is GWs + DT + $\Omega_1$ , and the magenta is GWs + DT + $\Omega_2$. On the plot below, the phase shift is measured. The green line is the wave shift generated by dynamical tides, cyan line is the wave shift generated by the spin $\Omega_1$, while the magenta line is the wave shift by  $\Omega_2$. }
\label{fig:tidesspin}
\end{figure}

Then we add the dimensionless spin, $ \chi =[0.1,0.2]$, to the NS in the same BH-NS system. We refer to the spin, $\Omega_{1}$ and $\Omega_{2}$, for the  dimensionless spin respectively, $\chi = [0.1, 0.2]$, that is obtained form the relation (\ref{eq:chi}). On the top figure of fig. \ref{fig:tidesspin}, we show the extracted gravitational waveform from simulations containing the different effects. The green waveform is GWs + dynamical tides (DT), the cyan is GWs + DT + $\Omega_1$ , and the magenta is GWs + DT + $\Omega_2$. On the plot below, the phase shift is measured. The green line is the wave shift generated by DT, the cyan line is the wave shift generated by the spin $\Omega_1$, while the magenta line is the wave shift due to $\Omega_2$. The wave shift generated by dynamical tides is the same as before. The spin, $\Omega_{1}$, generates a wave shift of $\Delta \varphi \simeq 0.45$, while $\Omega_{2}$, generates a wave shift of $\Delta \varphi \simeq 1.8$. Therefore, the corrections due to spin-size effects, can be the leading order contribution from the stellar structure. The spin evolution is negligible, having the same final value as the initial, which is expected from the aligned spin case.

\begin{figure}
\centering
    \includegraphics[width =0.9\textwidth]{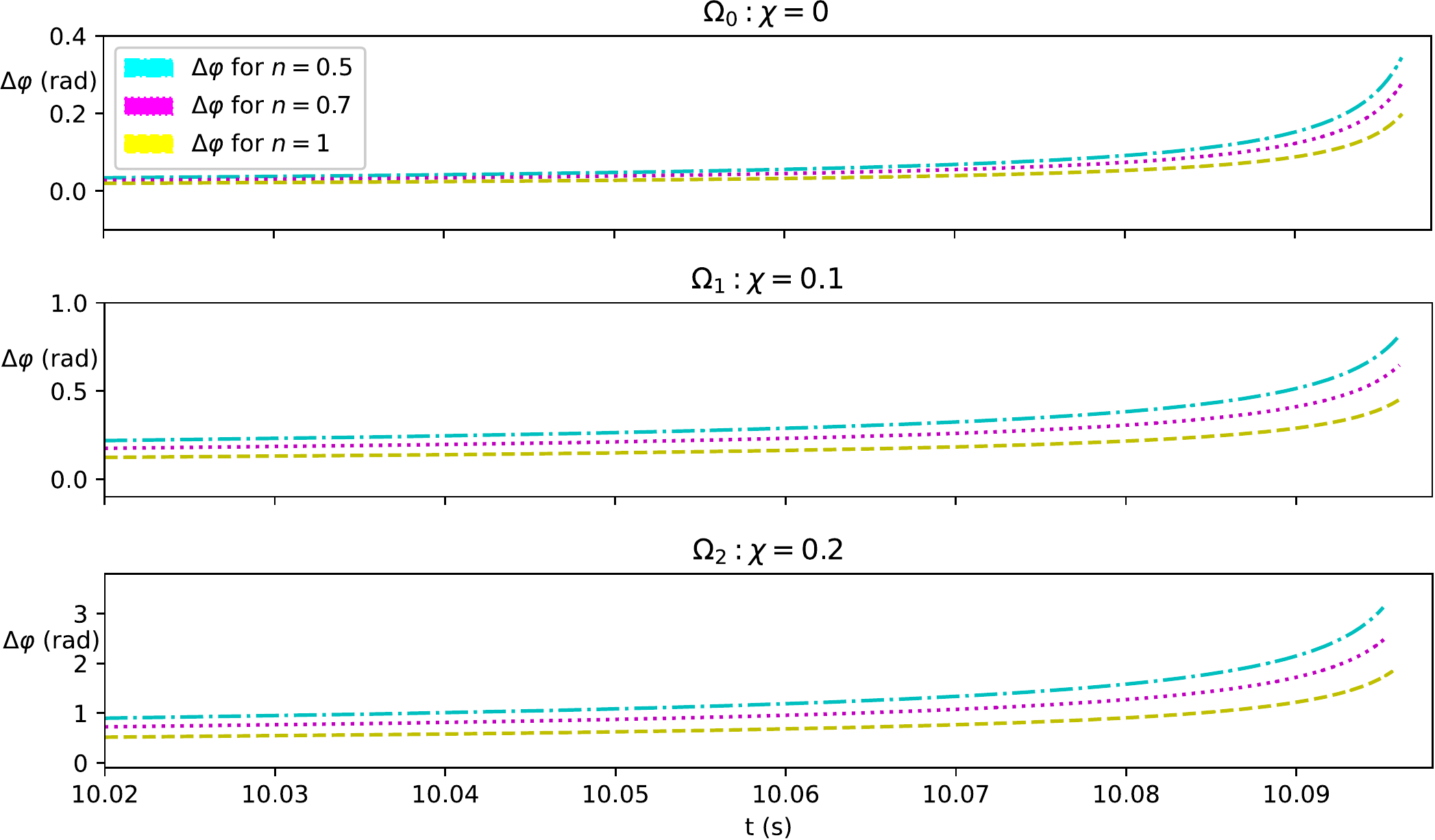}
\caption{Phase shifts from the late inspiral of the same BH-NS interaction described in figure \ref{fig:tides} and \ref{fig:tidesspin}, but with different dimensionless Love numbers, with corresponding polytropic index $n = [0.5,0.7,1]$. In all the figures, the line color refers to different polytropic indices for the NSs. On the top figure we show the different phase shift generated by static tides given different equations of state. The middle plot is the phase shift with respect to the top figure due to the inclusion of spin/size effects, for which we have added spin $\Omega_1$. On the plot below we show the phase difference from the top plot and the inclusion of the spin $\Omega_2$.}
\label{eq:tidesspineos}
\end{figure}

Finally, by considering static effects, as well as spin-size effects, in figure \ref{eq:tidesspineos}, we show the role of different Love coefficients in the waveform by measuring
the phase shifts from the late inspiral of the same BH-NS interaction described in figure \ref{fig:tides} and \ref{fig:tidesspin}, but with different dimensionless Love numbers, with corresponding polytropic index, $n$, for the NS \cite{Hinderer:2007mb}. In all the figures, the line color refers to different polytropic indices for the NSs. The cyan line for $n = 0.5$, the magenta for $n = 0.7$ and the yellow for $n = 1$. On the top figure we show the different phase shift generated by considering only static tides given different equations of state, without any spin, which generates a phase shift, $\Delta \varphi \simeq [0.346,0.278,0.200] $ radians, for each $q = [0.5,0.7,1]$, respectively. The middle plot is the phase shift with respect to the top figure due to the inclusion of spin/size effects, for which we have added spin $\Omega_1$. On the plot below we show the phase difference from the top plot and the inclusion of the spin $\Omega_2$. The spin $\Omega_{1}$ generates a wave shift of $\Delta \varphi \simeq [0.808,0.649,0.458] $ radians, while $\Omega_{2}$ generates a wave shift of $\Delta \varphi \simeq [3.136,2.555,1.851] $ radians, for each $n = [0.5,0.7,1]$. As expected, our results shows that the smaller the value of $n$, the more compact the star, thus the stronger the signatures in the signal.

\subsubsection*{Binary black hole coalescence}

\begin{figure}
\centering
    \includegraphics[width =\textwidth]{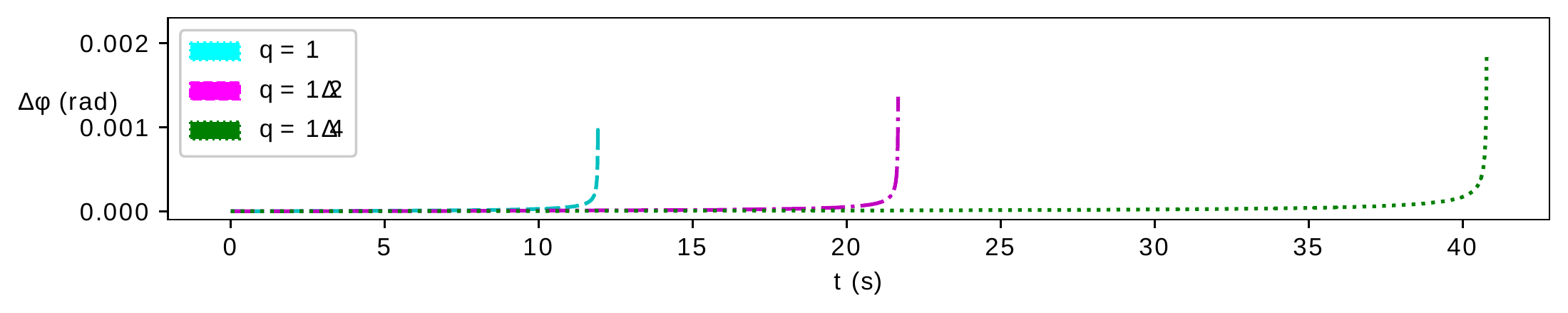}
\caption{The phase shift of the gravitational waveform from the late inspiral of three different BBHs that contains dissipative effects. We have performed three different simulations, each of them starting with initial $f_{gw} = 10$ Hz, but differing in the mass ratio, $q = m_2/m_1$, of the binary. The cyan line is for an equal mass case, $q = 1$, with $m_1 = 20 M_{\odot}$. The magenta line is for the unequal mass case, $q = 1/2$, and the green line for $q = 1/4$, both with $m_1 = 20 M_{\odot}$.}
\label{fig:dissipative}
\end{figure}

We have performed simulations for three different non-spinning BBHs to study the effects of dissipation in the waveform, which differ in the mass ratio of the binary, $q = m_2/m_1$. All of the simulations are set with an initial gravitational wave frequency, $f_{gw} = 10$ Hz. The cyan line is for an equal mass case, $q = 1$, with $m_1 = 20 M_{\odot}$. The magenta line is for the unequal mass case, $q = 1/2$, and the green line for $q = 1/4$, both with $m_1 = 20 M_{\odot}$. The phase shift due to dissipative effects is, $\Delta \varphi \simeq [0.0009, 0.0013, 0.0018]$, for each $q = [1,1/2,1/4]$ respectively. Our results shows that the effect of dissipation can increases for more unequal cases, but this also depends on the mass ratios taken into account. For instance, a simulation of the first detected binary black hole merger, GW150914, shows that dissipative effects makes a phase shift of $\Delta \varphi \simeq 0.0004$, which is the smallest phase shift of all. 

\subsubsection*{Neutron star binaries}

\begin{figure*}
    \centering
    \includegraphics[width=\textwidth]{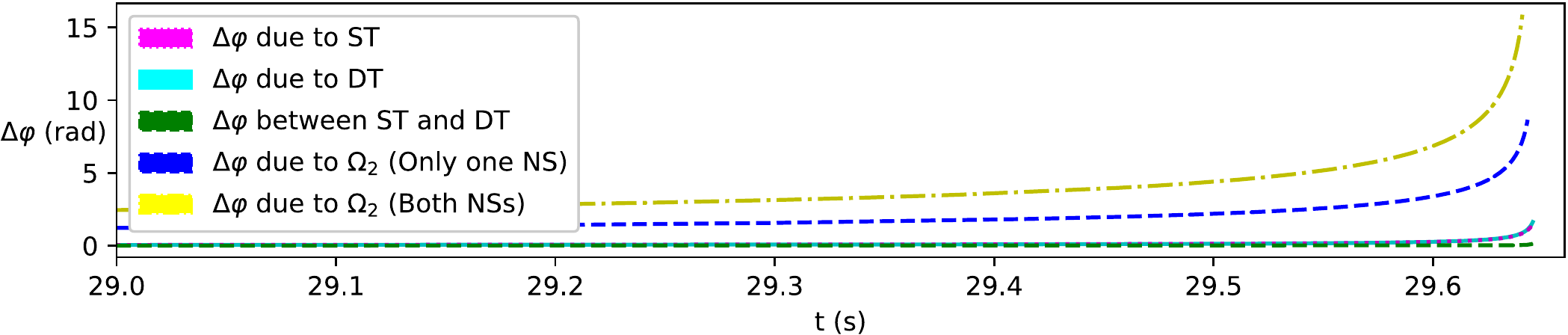}
    \caption{Phase shifts from simulations of a NS-NS coalescence with an initial frequency of $f_{gw}$ . We add GWs, DT and then spin to the NSs, first only to one member and then to both. The magenta color is the phase shift due to static tides, the cyan for dynamical tides, and the green the difference between both. The blue line is the phase shift due to one spinning star with $\chi = 0.2$, and the orange line for the case in which both stars are spinning with same magnitude and direction (aligned spins).  }
    \label{fig:NSNS}
\end{figure*}

We set an equal mass NS-NS binary with initial frequency of $f_{gw} = 40$ Hz. In order to illustrate different scenarios, we run four different simulations that includes GWs, dynamical tides and aligned spin, $\chi = 0.2$, to the NSs, first only to one member and then to both. In figure \ref{fig:NSNS}, we show that for tidal effects, the contribution to the wave shift due to the static term is of $\Delta \varphi \simeq 1$ radians, while the dynamic part contributes to the latter by $\Delta \varphi \simeq 0.1$ radians.  By comparing to the wave shift of the BH-NS case, we find that tidal effects in a NS-NS binary have a stronger effect on modifying the waveform. The role of spin-size effects is also enhanced in the NS-NS binary case, for which the waveform is shifted by, $\Delta \varphi \simeq 8 $ radians, with only one spinning star, and $\Delta \varphi \simeq 15 $ radians, from both spinning NS with same $\Omega_2$ aligned to the orbital momentum of the binary. Each of both spinning NS contributes the same phase shift, and just like in the BH-NS system, the spin evolution is negligible, having the same final value as initial. 

\section{Discussion}
\label{sec:discussion}

In this work we have implemented the numerical evolution of the late inspiral of compact objects within the EFT framework using a high accuracy 4th order Hermite integrator for point particle simulations. We have successfully included in the dynamics, leading order spin corrections due to gravity, dynamical tidal and dissipative effects, and the 2.5 PN corrections due to GW radiation. We have tested our implementation with analytical results, and matched the coefficients of the effective theory at each time step from the numerical simulations as an internal test. We have extracted the leading order gravitational waveform as a function of the orbital frequency of the binary, and showed the role of the coefficients by studying the overall phase evolution of the waveform due to the different effects. 

Although our current implementation is inaccurate for reproducing high order PN waveforms, what can be learned from our numerical experiments is the order at which the effects of the stellar structure arise in the gravitational waveform. We have found that in the BH-NS case, the coefficients due to static tides can be measured if LIGO-Virgo observatories, and future detectors, are sensitive enough to measure a shift in the waveform of, $\Delta \varphi \simeq O(10^{-1})$ radians, which will allow us to constrain between different Love coefficients. Furthermore, the stellar structure of spinning NS plays a role on shifting the waveform at order, $\Delta \varphi \simeq O (1)$ radians, which suggest that the spin-size effects, with coefficient $n_{\Omega}$, can be the leading order effect due to the stellar structure, depending on the value of the spin. On the other hand, to detect dynamical tides and better constrain the EOS of matter, it is necessary to measure, $\Delta \varphi \simeq O (10^{-3})$ radians, as well as for dissipative effects in the NS-BH case. 

In the case of the NS-NS binary, the wave shift due to tidal effects is enhanced, with the static tidal effects generating a shift on the wave of, $\Delta \varphi \simeq O (1)$ radians, while the dynamical tides, $\Delta \varphi \simeq O (10^{-1})$ radians, from the contribution of both stars. Just as in the case of the BH-NS, the gravitational spin correction is leading order, and could generate a wave shift of $\Delta \varphi \simeq O (10^{})$ radians, being the first coefficient to match from observations. Nevertheless, the dynamics of nonaligned spins with different magnitudes may complicate this task, as well as binaries with members described by different equations of state. Finally, from the BBH simulations including dissipation, we have shown that their effects generate a shift in the waveform at order,  $\Delta \varphi \simeq O (10^{-3})$ radians. 

Thus, we conclude that the first coefficient to match, given the order at which it contributes and which requires the least sensitivity, is the coefficient, $n_{\Omega}$, from the spin-size effects of spinning NSs, both in BH-NS and NS-NS binaries. Then the static tidal coefficient in BH-NS binaries, and the dynamical tides coefficient in NS-NS binaries. Finally dynamical tides in BH-NS interactions, and dissipative effects in both BH-NS and BBH, which play a role roughly at the same order, and which requires the most sensitivity for the observatories to measure.   We argue that the sensitivity required for the LIGO-Virgo and future GWs observatories to constrain the stellar structure of the compact objects with high precision, needs to measure wave shifts with an accuracy of at least, $\Delta \varphi \simeq O (10^{-4})$ radians.

This is the first step towards constructing realistic stellar models and extracting accurate GWs within the EFT framework for a wide class of astrophysical objects in the region in which the dynamics can be described by the post-Newtonian expansion. In this sense, this framework can allows us to create a rich template bank of waveforms for different classes of systems, such as the ones described in this work, and many others that can be detected by ground and space based GW detectors.

\acknowledgments

We are extremely grateful to R. Penco,  W. Goldberger, T. Hinderer, J. Steinhoff, L. Heisenberg, J. Samsing,  M. Levi, R. Yarza, and H. S. Chia for the many enlightening conversations. I.M. is particularly grateful to E. Ramirez-Ruiz, S. Rosswog and S. Nissanke for their support in early stages of the work. We gratefully acknowledge support from the University of Cape Town Vice Chancellor's Future Leaders 2030 Awards programme which has generously funded this research and support from  the South African Research Chairs Initiative of the Department of Science and Technology and the NRF. This research was supported in part by the National Science Foundation under Grant No. NSF PHY-1748958.


\bibliography{bibeft}

\end{document}